\documentclass[conference]{IEEEtran}

\IEEEoverridecommandlockouts

\usepackage{amsmath,amsfonts}
\usepackage{array}
\usepackage{textcomp}
\usepackage{stfloats}
\usepackage{url}
\usepackage{graphicx}
\usepackage{MnSymbol}
\usepackage{color}
\usepackage{comment}
\usepackage{acronym}
\usepackage{empheq}
\usepackage{enumitem}
\usepackage{cite}

\def\BibTeX{{\rm B\kern-.05em{\sc i\kern-.025em b}\kern-.08em
    T\kern-.1667em\lower.7ex\hbox{E}\kern-.125emX}}

\newcommand{\bs}[1]{\boldsymbol{#1}}

\acrodef{sime}[SIME]{Single Machine Equivalent}
\acrodef{omib}[OMIB]{One-Machine Infinite Bus}
\acrodef{coi}[COI]{Center of Inertia}
\acrodef{wscc}[WSCC]{Western System Coordinating 
Council}
\acrodef{eeac}[EEAC]{Extended Equal-Area Criterion}

\begin{document}

\title{On the Mechanism of Multi-Cluster System Separation Under Extreme Events\\%
\thanks{This work is supported by the Environmental Protection Agency~(EPA) by funding N.~Sofos and F.~Milano under project CRIE, Grant No.~2024-CE-1266; and by the European Commission and the Sustainable Energy Authority of Ireland, by funding G.~Tzounas under project NU-ACTIS, Grant Nos. 101069750 and 24/RDD/1390, respectively.}
}

\author{\IEEEauthorblockN{Nikolaos Sofos, Georgios Tzounas, and Federico Milano}
\IEEEauthorblockA{\textit{School of Electrical and Electronic Engineering} \\
\textit{University College Dublin, Ireland}\\
  nikolaos.sofos1@ucdconnect.ie, 
  georgios.tzounas@ucd.ie, 
  federico.milano@ucd.ie}
}

\maketitle

\begin{abstract}
  Historical blackout events demonstrate that, under complex multi-event contingencies, interconnected power systems can split into multiple areas exhibiting distinct dynamic behavior while remaining physically connected. This work provides a rationale for this phenomenon. Typical $N-1$ contingency analysis leads to a response dominated by a single unstable mode, supporting the two-cluster conjecture used by transient stability assessment tools such as the \ac{sime} method. However, extreme events, often involving cascading $N-k$ contingencies, can excite multiple unstable modes. We establish a proof of concept based on linear modal expansion to show how such concurrent excitation leads to multiple coherent areas.  The resulting behavior is illustrated through simulations on the standard \ac{wscc} 9-bus, IEEE 14-bus, and IEEE 39-bus systems, demonstrating the limitations of conventional two-cluster representations under extreme events.
\end{abstract}


\section{Introduction}
\label{sec:intro}

Extreme events, that is, multiple large disturbances that may lead to a blackout, have historically been rare but have become increasingly relevant in recent years.  Given their potentially severe consequences, the impact of such events on power systems has become the subject of intense study \cite{Panteli:2015, panteli2016power, bie2017battling, Perera:2020, gonccalves2024extreme, xu2024resilience}.  This paper focuses, in particular, on the formation of multiple coherent groups of generators following multiple contingencies.

Analyses of major blackout events show that they are typically caused by cascading failures \cite{dobson2007complex} and can involve the formation of several distinct areas before collapse.
For instance, during the August 10, 1996 Western North American blackout, growing oscillations led to uncontrolled system separation into multiple islands \cite{kosterev1999model}.
Similarly, the November 4, 2006 event in the Continental European grid resulted in its separation into three distinct areas \cite{li2007analysis}. 

Power system literature has extensively studied multi-group partitioning through slow-coherency analysis \cite{chow1982time}, controlled islanding schemes \cite{you2004slow}, and spectral graph clustering \cite{sanchez2014hierarchical}.  However, these frameworks focus primarily on identifying system partitions or designing controlled defense mechanisms.  The underlying physical mechanism linking $N-k$ contingencies to multi-cluster separation during extreme events remains largely unexplored.

On the other hand, major events associated with transient instability, namely, synchronous machines going out of step, are typically treated as a two-cluster phenomenon, despite the possibility that power systems may separate into several areas following large disturbances.  
The possibility of separation into more than two groups is neither discussed nor excluded by conventional definitions of transient stability \cite{Kundur:2004}.
However, direct and hybrid stability assessment methods 
are commonly formulated on the principle that loss of synchronism within a time frame of 3 to 5 s leads to the separation of machine trajectories into exactly 
two clusters \cite{pavella1998power}. 

The Extended Equal-Area Criterion (EEAC) \cite{xue1989extended, extendedeqarea1992, xue1993critical} 
and the \acf{sime} method \cite{zhang1997sime} 
are founded on the conjecture that, regardless of network scale or complexity, generators will separate into exactly two clusters: a critical cluster of accelerating machines and a remaining cluster of trailing machines.  This simple binary split is consistent with typical $N-1$ contingency analyses, where the unstable response is generally dominated by a single mode, enabling the reduction of a complex multi-machine system to an \ac{omib} equivalent \cite{zhang1998sime, ernst2002unified}. 

The contribution of this paper is the conjecture that \textit{multiple clusters can happen if a set of contingencies occurs in a time interval sufficiently small compared to the dynamics of the clusters}.  With this assumption, the rationale behind the two-cluster split assumption used in transient stability analysis is statistically justified because the time frame of transient instability is generally too short for multiple contingencies to occur.  
For slower phenomena, such as those driven by voltage or frequency instability, it is statistically more likely that multiple contingencies and protection events (e.g., cascading) happen during the evolution of the phenomena themselves, as observed in practice.  As a proof of concept, the paper provides a linear analysis based on modal expansion to show how the excitation of multiple unstable modes can give rise to more than two system clusters.

Case studies on the WSCC 9-bus, IEEE 14-bus, and IEEE 39-bus test systems illustrate the multiple-group separation mechanism across transient, small-signal, and frequency stability domains, respectively.  

\section{Conjecture on Multi-Cluster Separation}

As observed in real-world blackouts, when a severe event involving multiple contingencies that affect different locations simultaneously occurs in a power system, system dynamics can show a clustering into several coherent groups, before the physical network is separated into electrically disconnected regions. 

Here, \textit{simultaneous} does not have the literal meaning of \textit{at exactly the same time}, but rather within a time interval shorter than the characteristic time scale of the instability triggered by the contingencies.  For a long-term instability leading to, for example, voltage collapse, multiple contingencies can happen in a relatively long time interval and still be considered part of the same ``event.''  This is consistent with observations of multi-cluster separation during relatively slow instability phenomena in real-world blackouts.  For transient instability, on the other hand, multiple contingencies have to happen in a time frame of a few hundreds of milliseconds to cause the formation of multiple clusters, thus making the occurrence of this event quite unlikely.

In conventional transient stability assessment tools, security assessment relies heavily on single $N-1$ contingency analysis.  Such single-fault events typically excite a single dominant mode of instability, causing rotor angles to diverge along a single state-space direction and yielding a standard two-cluster separation.  In contrast, extreme events that can occur in real-world scenarios usually involve $N-k$ multi-contingencies, such as sequences of faults, line disconnections, and protection actions occurring across distinct geographic locations of the same system.  Rather than exciting a single mode, these distributed $N-k$ disturbances induce complex electromechanical swings among generators across multiple areas of the network.

In a classical nonlinear power system model, the active power transfer between connected machines $i$ and $j$ is governed by network transfer capacity and is proportional to $\sin(\delta_i - \delta_j)$, with synchronizing torque coefficients proportional to $\cos(\delta_i - \delta_j)$. Under extreme $N-k$ contingencies, large local power imbalances between mechanical input and electrical output develop simultaneously in different parts of the grid. As rotor angle deviations increase, nonlinear power transfer boundaries saturate at multiple network interfaces. Consequently, competing synchronizing torques act on intermediate generators from different directions, preventing them from joining a single accelerating or decelerating group and resulting in multi-group separation.

\subsection{Proof of Concept}
\label{sec:proof_of_concept}

To illustrate how a multi-machine system can split into more than two groups, we consider the phenomenon for linear systems.  This section serves as a proof-of-concept showing how the simultaneous excitation of multiple unstable modes can lead to multi-group separation.  The analysis provides insight into the underlying mechanism, which is subsequently examined through simulations on fully-fledged nonlinear power system models in the case studies discussed in the next section.

Consider a linear system in its state-space form $\dot{\bs{x}} = \mathbf{A}\bs{x}$. Its trajectory is defined by the modal expansion:
\begin{equation}
    x_i(t) = \sum_{j=1}^{n} c_j e^{\lambda_j t} {r}_{i,j} \, ,
\end{equation} 
where $\lambda_j$ is the $j$-th eigenvalue of $\mathbf{A}$, with corresponding right and left eigenvectors $\bs{r}_j$ and $\bs{w}_j$, respectively, and $\mathbf{A}$ is assumed to be diagonalizable. The $i$-th element of $\bs{r}_j$ is ${r}_{i,j}$.
The projection of the initial conditions onto the $j$-th mode is $c_j = \bs{w}_j \mathbf{x}(0)$.

If the system is unstable and its response is dominated by a single unstable mode associated with a real eigenvalue $\lambda_1>0$, while all remaining modes are stable, the system diverges along the direction of $\bs{r}_1$.  For any two states $x_k, x_l$,  after the stable modes have sufficiently decayed:
\begin{equation}
    x_k(t) \approx \frac{{r}_{k,1}}{{r}_{l,1}} x_l(t) \, .
\end{equation}
Thus, the relative motion of the states is constrained by a single eigenvector, leading to a standard two-group separation based on the signs of its components.

If, instead, the system has two real unstable modes associated with $\lambda_1,\lambda_2>0$, while all remaining modes are stable, the system diverges along the two-dimensional subspace spanned by $\{\bs{r}_1, \bs{r}_2\}$.  For any two states $x_k, x_l$, after the stable modes have sufficiently decayed:
\begin{equation}
x_k(t) \approx
\frac{c_1{r}_{k,1}e^{(\lambda_1-\lambda_2)t} +
c_2{r}_{k,2}}{c_1{r}_{l,1}e^{(\lambda_1-\lambda_2)t}+c_2{r}_{l,2}} x_l(t) \, .
\end{equation}
Thus, under the simultaneous excitation of two unstable modes, the states can develop into multiple clusters. This behavior persists longer when the corresponding growth rates are closer, i.e., when $|\lambda_1-\lambda_2|$ is smaller.

\begin{figure}[htb!]
  \centering
  \includegraphics[width=0.99\linewidth]{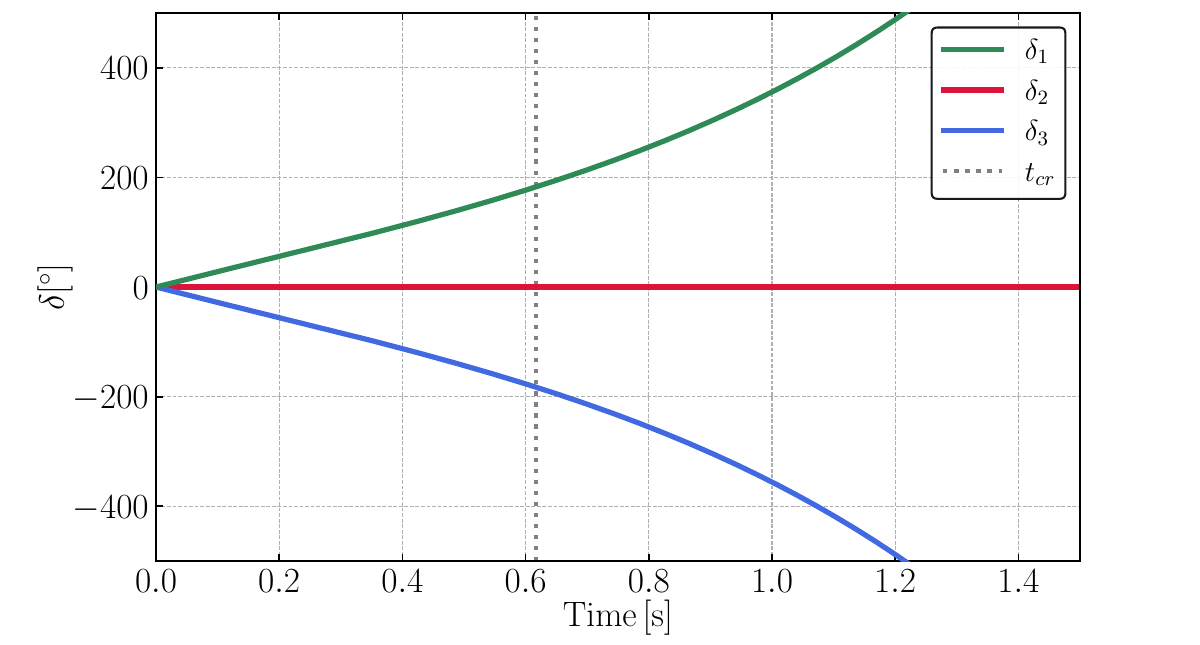}
  \vspace{-2mm}
  \caption{Linearized three-machine system: three-group rotor-angle separation.}
  \label{fig:linear_system}
\end{figure}

For illustration purposes, consider a three-machine system in which each synchronous machine is represented by the classical electromechanical model, i.e., by its swing equation, linearized around an equilibrium point. The system is symmetric with respect to G2. Generators G1 and G3 have identical inertia and damping constants and are connected to G2 through identical network branches, so that G2 is located electrically in the middle of the system. The equilibrium point is selected such that the state matrix has two real positive eigenvalues. Figure~\ref{fig:linear_system} shows the rotor angle deviations of the three machines following a small perturbation of the initial state.

Let $t_{cr}$ denote the time at which any two machines reach a mutual angular deviation of 360°, which occurs at approximately 0.6 s. At $t_{cr}$, the rotor angles of G1 and G3 have diverged in opposite directions from their initial values, whereas the rotor angle of G2 remains at 0° throughout the simulation. This occurs because the system is perfectly symmetric, causing the opposing forces acting on G2 from the other two generators to cancel each other exactly.  Thus, partitioning the generators into two groups would fail to capture the actual dynamic response of the system, as all three generators exhibit distinct dynamic behaviors.

\section{Examples}
\label{sec:examples}

The examples presented below showcase the different ways in which a power system can split into multiple groups.  Three power systems are discussed, each one related to a different type of stability analysis, namely, transient stability using the \ac{wscc} 9-bus system; small-signal stability using the IEEE 14-bus system; and frequency stability using the IEEE 39-bus system. Simulations were carried out using the Python-based software tool Dome~\cite{milano2013python}.

\subsection{Transient Stability - WSCC 9-bus System}

This example illustrates how the angles of the generators of a power system can form multiple groups when the system becomes unstable following an extreme disturbance. The \ac{wscc} system features three generators: G1, G2, and G3, located at buses 1, 2 and 3, respectively. 

In this study, the extreme event is modeled as a spatially distributed double contingency. At $t = 0$~s, a three-phase fault occurs at bus~2 and is subsequently cleared at $t = 0.16$~s. At $t = 0.15$~s, an additional load of $500$~MW is connected to bus~3 for a duration of $0.15$~s, before being disconnected at $t = 0.3$~s. 
While a simultaneous fault and massive load connection represents an extraordinarily severe scenario, this $N-k$ contingency effectively mimics the competing forces present during cascading events, causing multi-group separation.

\begin{figure}[htb!]
  \centering
  \includegraphics[width=0.99\linewidth]{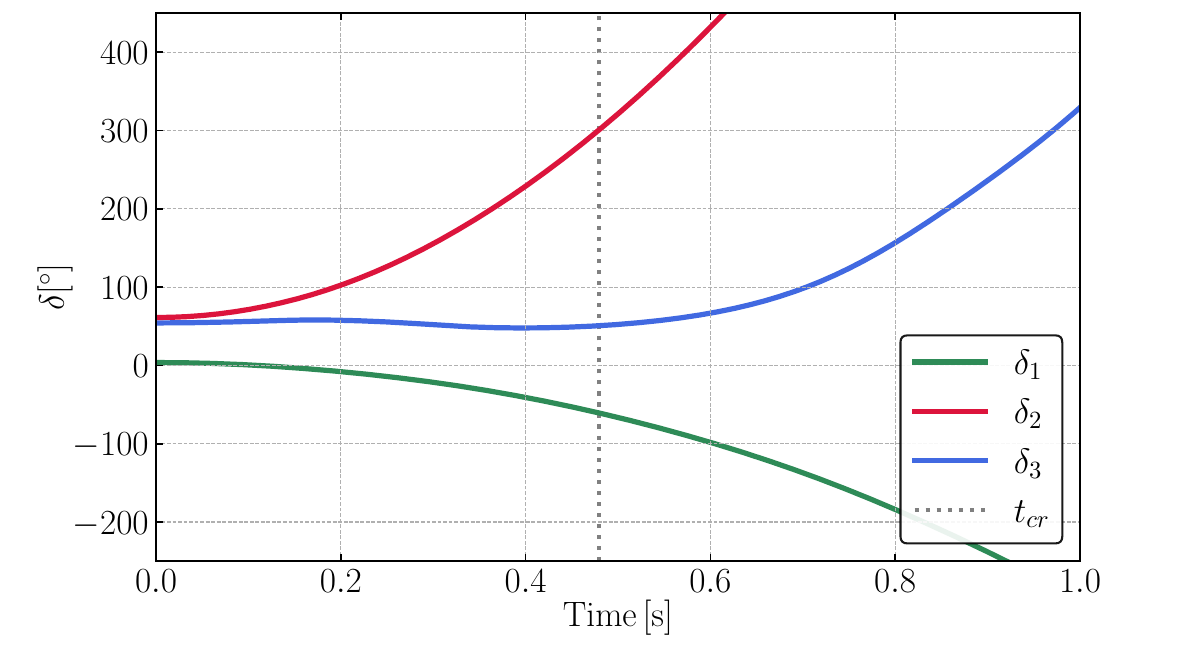}
  \vspace{-2mm}
  \caption{WSCC 9-bus system: three-group separation.}
  \label{fig:wscc_extreme_event_1}
\end{figure}

The three-phase fault induces a rapid acceleration in G2, while the sudden load connection forces G1 to decelerate. Thus, G3 is subjected to competing, opposing synchronizing torques from both the accelerating and decelerating machines. This causes the rotor angle of G3 to remain bounded near its pre-fault equilibrium value for a transient window, making it difficult to cluster it with either of the diverging generators. This behavior is illustrated in Fig.~\ref{fig:wscc_extreme_event_1}. 

The response of the rotor angles in Fig.~\ref{fig:wscc_extreme_event_1} shows that at $t_{cr}$, the rotor angle of G3 remains near its pre-fault value, while the rotor angles of G1 and G2 diverge in opposite directions. Thus, the reduction of the system into an \ac{omib} would fail to capture the system dynamics at $t_{cr}$.

For $t > t_{cr}$, the angle of G3 subsequently diverges from that of G1.  Following the steps of the assessment methods mentioned in the introduction, such as the \ac{sime} method, would have put these two generators in the same group. This delayed divergence occurs due to the interaction between G3 and G1, which temporarily cancels the accelerating torque of G3. 

\begin{figure}[htb!]
\centering
\includegraphics[width=0.99\linewidth]{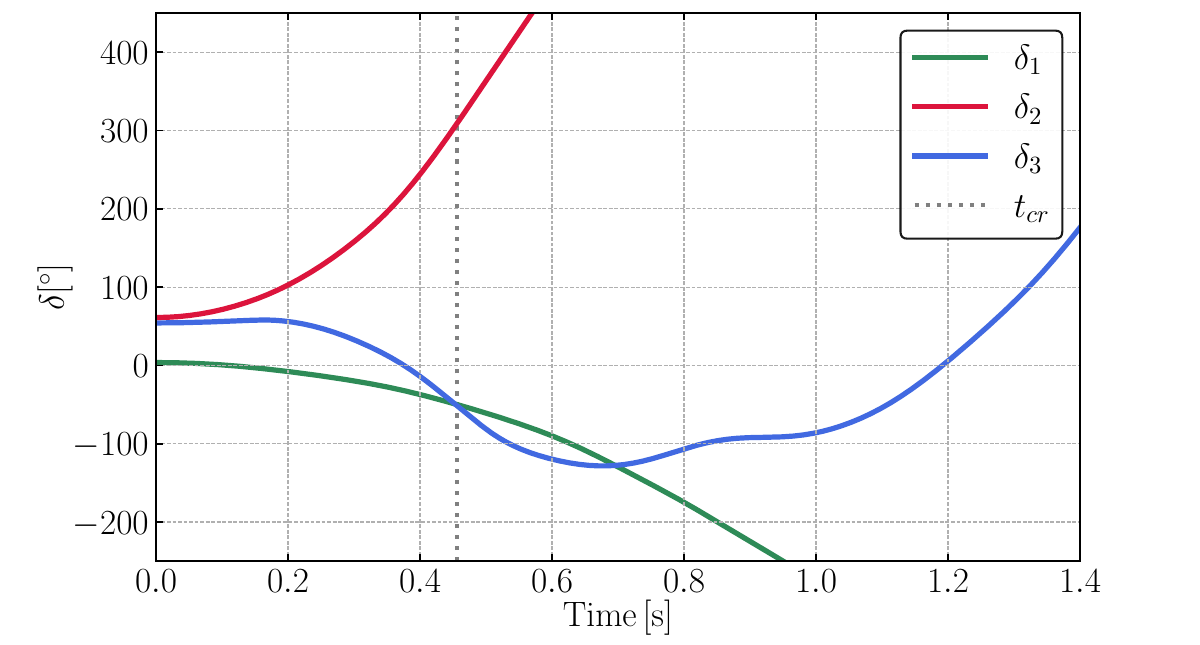}
\vspace{-2mm}
\caption{WSCC 9-bus system: three-group separation.}
\label{fig:wscc_extreme_event_2}
\end{figure}

To explore how the disturbance parameters influence the trajectories of the rotor angles of the machines, we change the disturbance durations: the three-phase fault is cleared at $t = 0.300$~s, and the load connection is sustained for a duration of $0.350$~s. 
As shown in Fig.~\ref{fig:wscc_extreme_event_2}, G3 initially exhibits a response similar to G1. However, for $t > t_{cr}$, the moment when standard transient stability tools would typically terminate the simulation to form the clusters of generators, the rotor angles of G3 and G1 actually diverge.

\begin{figure}[htb!]
\centering
\includegraphics[width=0.99\linewidth]{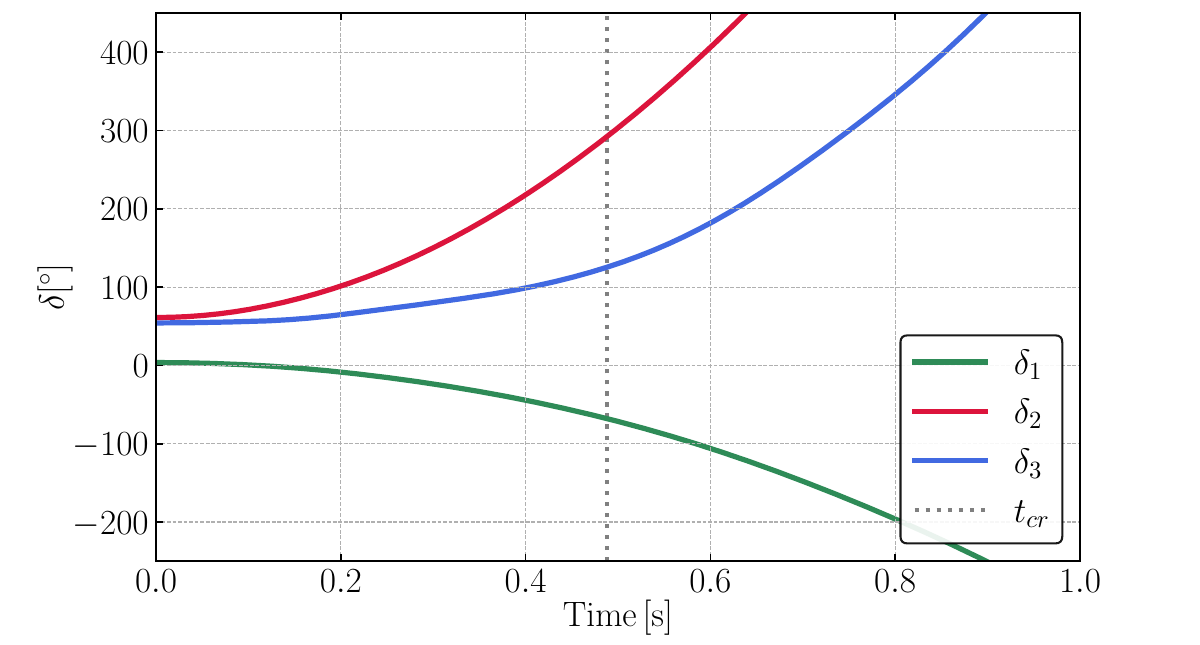}
\vspace{-2mm}
\caption{WSCC 9-bus system: two-group separation.}
\label{fig:wscc_extreme_event_3}
\end{figure}

Finally, to demonstrate the importance of the timing of the disturbances, we reduce the fault duration so that the fault is cleared before the load is connected and the two events do not overlap. We also delay the load connection further in time. Specifically, the three-phase fault is cleared at $t=0.107$~s, while the load is connected for $0.15$~s, from $t=0.200$~s to $t=0.350$~s. As shown in Fig. \ref{fig:wscc_extreme_event_3}, in this case the load connection is not sufficient to cause the formation of three groups. At $t_{cr}$, only two groups have formed: one consisting of G2 and G3, and the other consisting of G1.

This example illustrates the physical mechanism of multi-group angle separation following an extreme disturbance. The generator rotor angles show how opposing synchronizing forces can cause an intermediate generator to form a third cluster, demonstrating that the two-cluster separation conjecture is not universally valid under severe multi-event contingencies.

\subsection{Small-Signal Stability - IEEE 14-bus System}

The IEEE 14-bus system is used to demonstrate how the simultaneous excitation of multiple unstable modes can lead to multi-group system separation. Unlike transient stability analysis, which typically involves structural switching events, small-signal stability analysis shows that this multi-group splitting can also arise under small perturbations.

In synchronous generators, Automatic Voltage Regulators (AVRs) are utilized to maintain terminal voltage stability. However, excessively high AVR amplifier gains ($K_a$) can have the opposite effect. In this case study, the AVR gains of the generators are set to $K_{a,G1} = 100$, $K_{a,G2} = 1500$, $K_{a,G3} = 1500$, and $K_{a,G4} = 1500$. 

\begin{figure}[htb]
  \centering
  \includegraphics[width=0.95\linewidth]{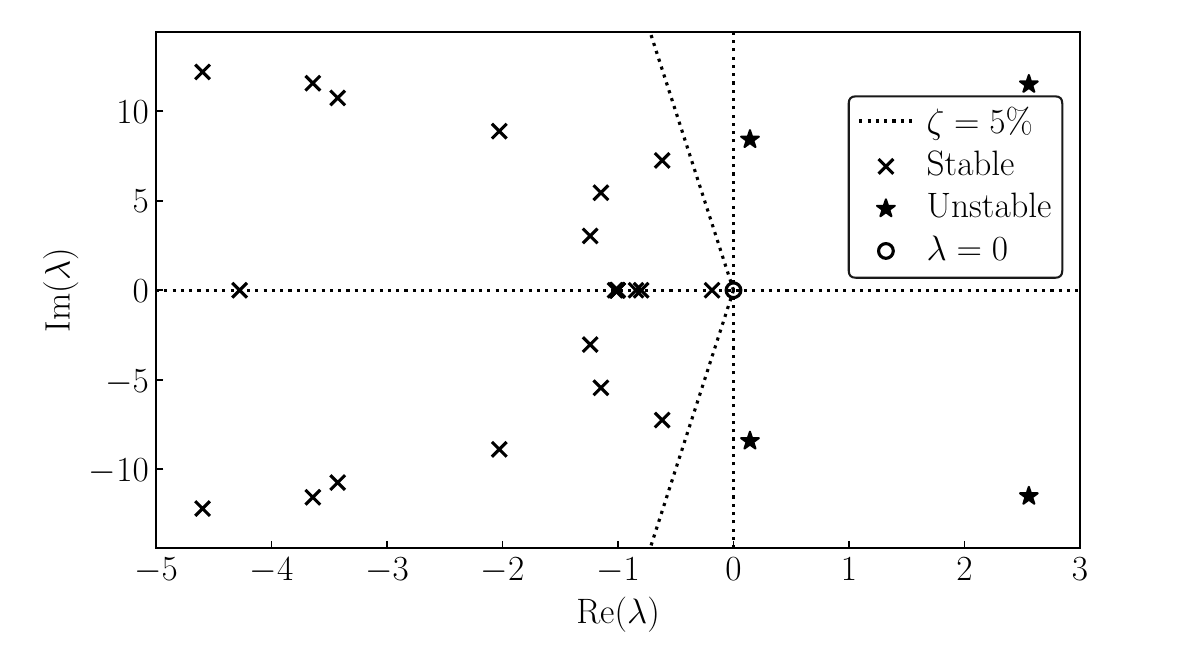}
  \vspace{-2mm}
  \caption{IEEE 14-bus system: Eigenvalues.}
  \label{fig:14_Bus_eigs}
\end{figure}

Due to the simultaneous tuning of high gains across all system units, two distinct complex-conjugate eigenvalue pairs cross the imaginary axis into the right-half complex plane ($\operatorname{Re}(\lambda) > 0$), as illustrated in Fig.~\ref{fig:14_Bus_eigs}. 

This example illustrates that the multi-group separation phenomenon can also be observed through small-signal stability analysis. By demonstrating that a power system model can contain multiple positive real-part eigenvalues, this scenario validates the modal analysis in Section~\ref{sec:proof_of_concept}, showing that multi-mode instability is a primary mechanism driving system separation into more than two groups during an extreme event leading to a blackout.

\subsection{Frequency Stability - IEEE 39-bus System}

The IEEE 39-bus system is used to illustrate how generator frequencies can form multiple groups when the system becomes unstable following an extreme disturbance, providing insight into the mechanism behind the 2006 European blackout, where the interconnected system split into three areas with different frequencies.

\begin{figure}[htb]
  \centering
  \includegraphics[width=0.9\linewidth]{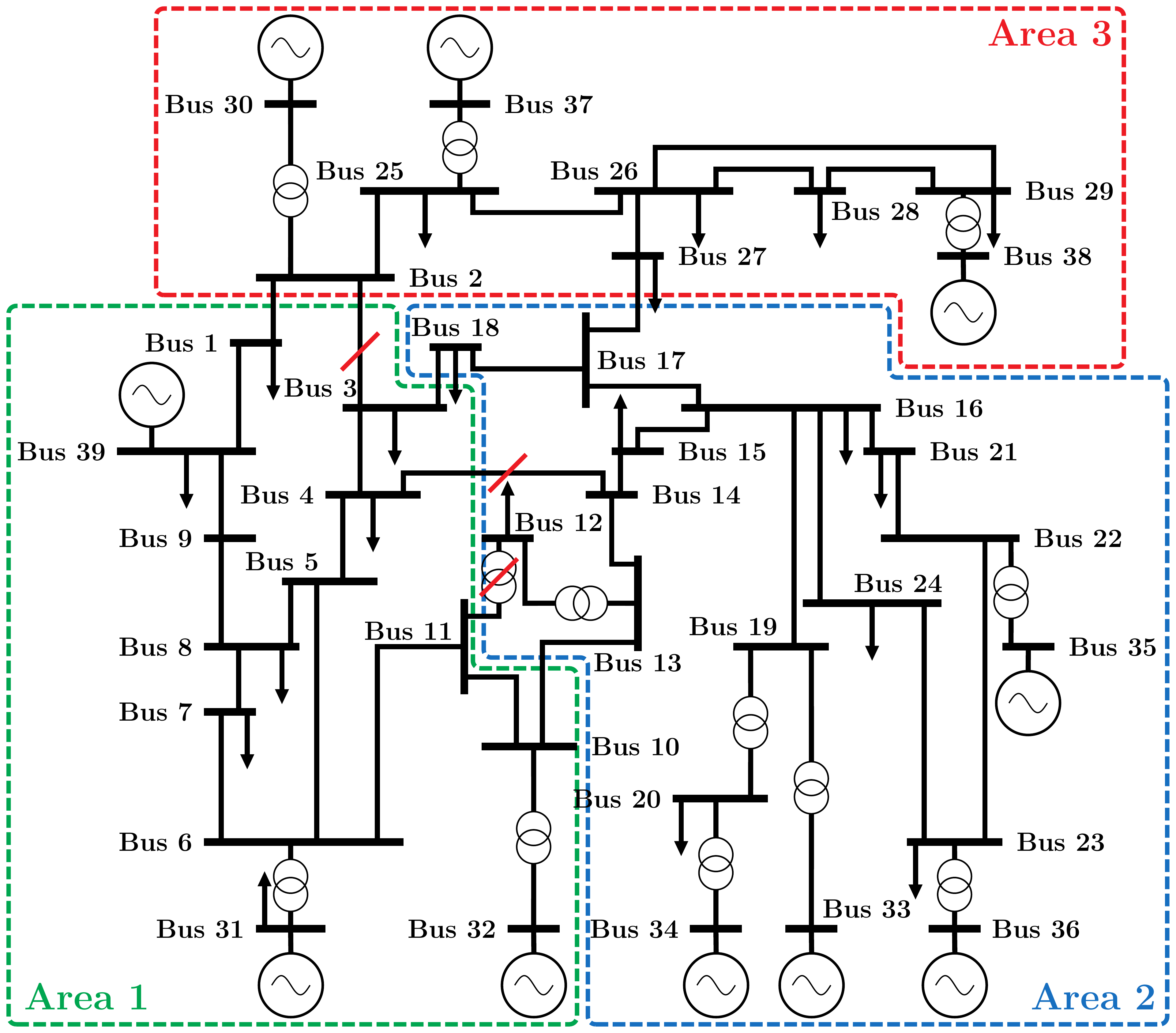}
  \vspace{-2mm}
  \caption{IEEE 39-bus system: single-line diagram.}
  \label{fig:39_Bus}
\end{figure}

The modified IEEE 39-bus system that is used in the examples is shown in Fig. \ref{fig:39_Bus}. The figure also shows the three areas into which the system separates following the sequence of disturbances considered in this example. The red diagonal lines indicate the transmission lines that are disconnected during the disturbance sequence.

To create the conditions under which the generators separate into these three areas, the system was modified. Specifically, the active power output of the generators was reduced, and the lengths of the transmission lines connecting buses 1 and 2, 17 and 27, 3 and 17, and 10 and 13 were increased by a factor 1000. 
This was done to increase the electrical separation between the three areas after the sequence of line disconnections, while maintaining the physical connectivity of the network.

The sequence of events is as follows. First, the transmission line connecting buses 2 and 3 is disconnected at $t=1$ s.  Next, the transmission line connecting buses 4 and 14 is disconnected at $t=1.2$ s.  Finally, the transformer connecting buses 11 and 12 is disconnected at $t=1.5$ s.

\begin{figure}[t!]
\centering
\includegraphics[width=0.95\linewidth]{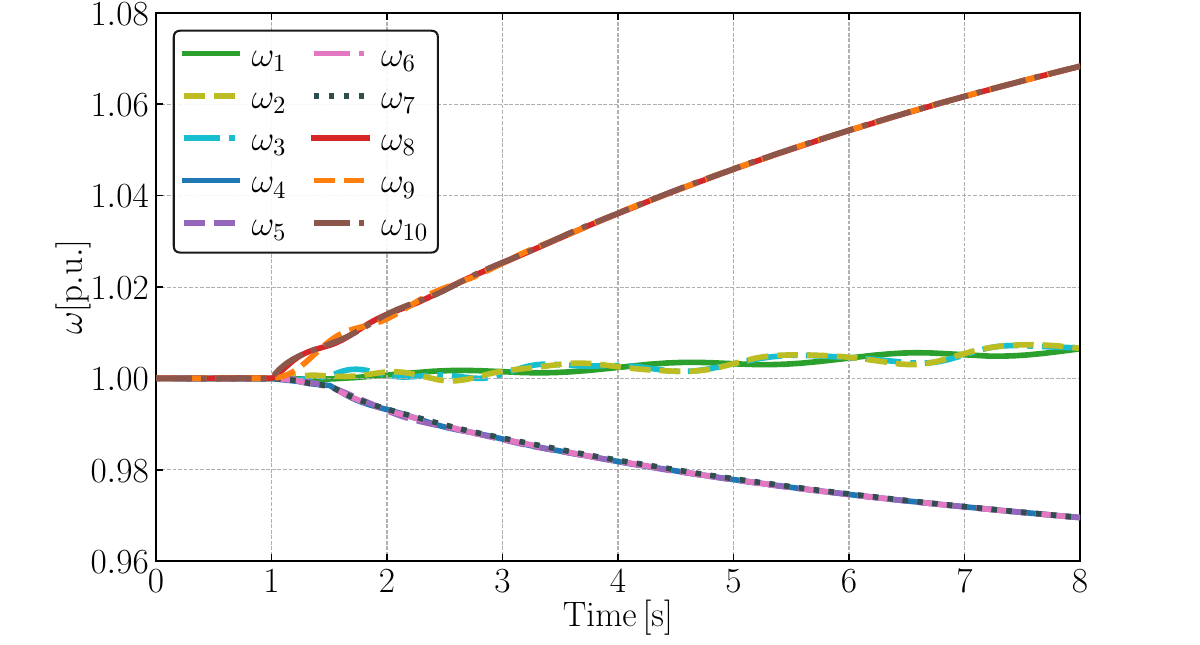}
\caption{IEEE 39-bus system: three-group machine separation.}
\label{fig:ieee_39_extreme_event}
\vspace{-4mm}
\end{figure}

Figure \ref{fig:ieee_39_extreme_event} shows the rotor speeds of all the generators in the system.  The sequence of contingencies causes the generators to separate into three clusters, each operating at a different frequency, before the system eventually becomes unstable.

The generators in area 1 form the first cluster, with their frequencies remaining between those of the other two clusters.  Cluster 1 consists of generators 1, 2, and 3, connected to buses 39, 31, and 32, respectively.  The generators in Area 2 form the second cluster, with their frequencies decreasing below the nominal frequency.  Cluster 2 consists of generators 4, 5, 6, and 7, connected to buses 33, 34, 35, and 36, respectively.  Finally, the generators in Area 3 form the third cluster, with their frequencies increasing above the nominal frequency.  Cluster 3 consists of generators 8, 9, and 10, connected to buses 37, 38 and 30, respectively.

This frequency divergence is driven by local active power imbalances coupled with restricted inter-area transfer capacity.  Following the sequence of line trippings, Area~3 retains a net generation surplus ($P_G > P_L$).  Because the increased electrical distance of the remaining interconnecting lines constrains active power transfer to neighboring zones, this excess mechanical power accelerates the local synchronous rotors, driving the frequency upward.  Conversely, Area~2 suffers a net power deficit ($P_G < P_L$), causing its generators to decelerate.  Area~1 maintains a relative balance between local generation and load, causing its frequency trajectory to remain bounded between the accelerating and decelerating clusters. 

This example demonstrates that, under a sequence of events, an interconnected system can separate into multiple asynchronous frequency zones prior to collapse.  These dynamics cannot be captured by a binary separation assumption, as each isolated or loosely coupled area follows an independent frequency trajectory dictated by its local power balance.

\section{Conclusions}
\label{sec:conclusion}

This paper discusses the underlying mechanisms of multi-group system separation through an analytical rationale, supported by case studies on the WSCC 9-bus, IEEE 14-bus, and IEEE 39-bus systems.  This phenomenon, which is frequently observed during real-world cascading blackouts, was illustrated across the domains of transient, small-signal, and frequency stability. While the presented examples highlight the limitations of stability assessment methods that rely on binary, two-cluster assumptions, such as the \ac{omib} equivalent, the primary contribution of this work is establishing the conditions that trigger multi-area system separation.  By linking $N-k$ contingencies to the time scale of the instability, the paper shows how such events can excite multiple unstable modes and provides an analytical and physical rationale for multi-cluster separation during extreme blackout events.


\end{document}